\documentstyle[prd,aps]{revtex}

\def\slash {\put(1,1){/}}
\def\backabit {\hskip -1.5pt}
\def\hdots#1(#2,#3)#4{\multiput(#2,#3)(#1 3,0){#4}{\circle*{2}}}
\def\Xdots(#1,#2){
   \multiput( #1, #1)( 2.12, 2.12){#2}{\circle*{2}}
   \multiput( #1,-#1)( 2.12,-2.12){#2}{\circle*{2}}
   \multiput(-#1, #1)(-2.12, 2.12){#2}{\circle*{2}}
   \multiput(-#1,-#1)(-2.12,-2.12){#2}{\circle*{2}}
}
\def\bigcirc (#1,#2,#3){\put(#1,#2){\circle{#3}}}
\def\fcircle (#1,#2,#3){\put(#1,#2){\circle*{#3}}}
\def\hcircles(#1,#2)#3{\multiput(#1,#2)(5,0){#3}{\circle{3}}}

\def\Floop{\hdots-(-16,0)4{\thicklines\bigcirc (0,0,28)}\hdots+(16,0)4}
\def\FloopX{{\thicklines\bigcirc (0,0,28)}\Xdots(12,4)}

\def\Bloop{ 
\bigcirc (-9, 9,3)\bigcirc (-5, 12,3)\bigcirc (0, 13,3)
                  \bigcirc ( 5, 12,3)\bigcirc ( 9, 9,3)
\bigcirc (-9,-9,3)\bigcirc (-5,-12,3)\bigcirc (0,-13,3)
                  \bigcirc ( 5,-12,3)\bigcirc ( 9,-9,3) }

\def\vertex#1{\bigcirc (#1 13,0,3)\bigcirc (#1 12,5,3)
              \bigcirc (#1 12,-5,3)\hdots#1(#1 17,0)4}
\def\loop#1{{\thicklines\bigcirc (#1 13,0,12)}\hdots#1(#1 22,0)3}
\def\BloopVV{\Bloop\vertex-\vertex+}
\def\BloopVL{\Bloop\vertex-\loop+}
\def\BloopLL{\Bloop\loop-\loop+}

\def\obdots#1#2{\multiput(#1 16.5,#2 2.5)(#1 1.5,#2 2.5){6}{\circle*{2}}}
\def\obdotl#1#2{\multiput(#1 19.5,#2 5.5)(#1 1.5,#2 2.5){4}{\circle*{2}}}
\def\verteX#1{\bigcirc (#1 14,0,3)\bigcirc (#1 12,5,3)
              \bigcirc (#1 12,-5,3)\obdots#1+\obdots#1-}
\def\loopX#1{{\thicklines\bigcirc (#1 14,0,12)}\obdotl#1+\obdotl#1-}
\def\BloopVVX{\Bloop\verteX-\verteX+}
\def\BloopVLX{\Bloop\verteX-\loopX+}
\def\BloopLLX{\Bloop\loopX-\loopX+}

\def\FPloop{ \hdots-(-16,0)4\hdots+(16,0)4
\fcircle (-12, 5,1)\fcircle (-9, 9,1)\fcircle (-5, 12,1)\fcircle (0, 13,1)
\fcircle (-12,-5,1)\fcircle (-9,-9,1)\fcircle (-5,-12,1)\fcircle (0,-13,1)
\fcircle ( 12, 5,1)\fcircle ( 9, 9,1)\fcircle ( 5, 12,1)\fcircle ( 13,0,1)
\fcircle ( 12,-5,1)\fcircle ( 9,-9,1)\fcircle ( 5,-12,1)\fcircle (-13,0,1)  }

\begin{document}

\hfill {\bf SMC-PHYS-155}

\begin{center}

{\large\bf Compositeness Condition}\footnote{
Invited talk given at Shizuoka Workshop on Masses and Mixing
of Quarks and Leptons, Shizuoka, Japan, March 1997.
To be published in Proceedings.
} \\[.2in]

{\bf  Keiichi Akama}\footnote{
\noindent E-mail: akama@tanashi.kek.jp, akama@saitama-med.ac.jp } \\

Department of Physics, Saitama Medical College\\
                     Kawakado, Moroyama, Saitama, 350-04, Japan \\

\end{center}

\leftskip 15mm\rightskip 15mm
{\abstract{
\baselineskip  = 10pt \small 
By solving the compositeness condition, 
	under which the Yukawa-type model coincides the NJL type model,
	we obtain the expressions for the effective coupling constants
	in terms of the compositeness scale 
	at the next-to-leading order in $1/N$.
In the NJL model with a scalar composite,
	the next to leading contributions are too large for $N_c=3$.
In the induced gauge theory with abelian gauge symmetry,
	the correction term is reasonably suppressed,
	while in the non-abelian gauge theory
	the corrections are suppressed only
	when it is asymptotically non-free.
}}

\leftskip 0cm\rightskip 0cm

\vglue.2in

\vskip 3mm\noindent 
{\large\bf 1. Introduction}

This talk is based on the recent works done in collaboration with
Takashi Hattori from Kanagawa Dental College.
The origin of the generations is one of the most challenging problem 
	in physics today and future.
Among various ideas, compositeness may be a natural strong candidate 
	for the solution of the problem.
Repeated appearance of the color triplets in the quark lepton spectrum
	seems to suggest existence of the common subconstituent 
	$c_i(i=1,2,3)$ which carries color \cite{PS},
and repeated appearance of the weak doublets 
	suggests existence of the common subconstituent 
	$w_i (i=1,2)$ which carries weak isospin \cite{AT}.
Then the quarks $q$ and leptons $l$ are composed of them as
\begin{eqnarray} 
	q\sim whc,\ \ \ \ l\sim wh', 
\end{eqnarray} 
where $h$ and $h'$ are subconstituents or sets of subconstituents
	(including empty sets) depending on details of models.
The weak boson $W_\mu ^i$ and the Higgs scalar $\phi $ can also be composites 
	like  
\begin{eqnarray} 
W_\mu ^i\sim \overline w_{\rm L}\tau ^i\gamma _\mu w_{\rm L} 
\ \ \ {\rm or} \ \ \ 
\overline q_{\rm L}\tau ^i\gamma _\mu q_{\rm L}
\ \  \  \ \ \ \ \  \ \ \ \ \ \ 
\phi \sim     \overline w_{\rm R} w_{\rm L}
\ \ \ {\rm or} \ \ \ 
\overline q_{\rm R} q_{\rm L}
\end{eqnarray} 
About twenty years ago, 
	we considered a composite model 
	of the Nambu-Jona-Lasinio type \cite{NJL},
	where the gauge bosons and the Higgs scalar 
	appear as composites of the quarks and leptons, 
	or subquarks \cite{AT,SS}.
In this type of model, the compositeness condition \cite{cc}
	plays an important role.
This is the condition that 
\begin{eqnarray} 
	Z_3=0
\end{eqnarray} 
	where $Z_3$ is the wave-function renormalization constant 
	of a boson in a Yukawa-type model.
Under this condition, the Yukawa type model with the elementary boson 
	becomes equivalent to the Nambu-Jona-Lasinio type model \cite{LM},
	and the elementary boson becomes a composite.
The compositeness condition imposes relations among 
	the coupling constants, masses 	and the compositeness scale.
When it is applied to the standard model 
	where gauge bosons and Higgs scalar 
	are taken as composites of the quarks and leptons,
	the relations indicate that at least one of the quarks should 
	have a mass of the order of the weak interaction scale \cite{AT}.
It looked puzzling because the known quarks at that time 
 	were much lighter than the weak scale.
Today, however, we know that 
 	the top quark has the mass of the order of the weak scale \cite{top},
 	and the relation derived from the compositeness condition
	becomes rather natural.
This fact called the revived attentions to the NJL-type model 
	of the spontaneously broken electroweak symmetry \cite{sb}.
Numerically, however, it does not precisely hold.
We need to consider how to make it more precise
	beyond the leading approximation in ${1/N}$ \cite{1/N}.

\vskip 3mm\noindent 
{\large\bf 2. Nambu-Jona-Lasinio Model}

\noindent {\bf 2.1 Compositeness Condition}

We consider the NJL model for the fermion 
	$\psi =\{\psi _1,\psi _2,\cdots ,\psi _N\}$ with $N$ colors
	given by the Lagrangian 
\begin{eqnarray} 
  {\cal L}_{\rm NJL}=\overline \psi i\slash \partial \psi  
                    + f|\overline \psi _{\rm L}\psi _{\rm R}|^2     
\end{eqnarray} 
	with $U(1)\times U(1)$ chiral symmetry.
In 3+1 dimensions, it is not renormalizable,
	and we assume a very large but finite momentum cutoff.
This Lagrangian ${\cal L}_{\rm NJL}$ is known to be equivalent 
 	to the linearized Lagrangian \cite{aux}
\begin{eqnarray} 
  {\cal L}'_{\rm NJL}
          =\overline \psi i\slash \partial \psi 
          +(\overline \psi _{\rm L} \phi  \psi _{\rm R}+{\rm h.c.})
          -{1\over f}|\phi |^2
\end{eqnarray} 
	which is written in terms of the auxiliary field $\phi $.
Now compare it with this Lagrangian of the renormalized Yukawa model,
\begin{eqnarray} 
  {\cal L}_{\rm Yukawa}&
     =&Z_\psi \overline \psi _{\rm r}i\slash \partial \psi _{\rm r}
     +Z_g g_{\rm r}(\overline \psi _{\rm rL} \phi _{\rm r} \psi _{\rm rR}
                         +{\rm h.c.})
\cr &&
     +Z_\phi |\partial _\mu \phi _{\rm r}|^2  
     -Z_\mu   \mu _{\rm r}^2|\phi _{\rm r}|^2 
     -Z_\lambda \lambda _{\rm r}|\phi _{\rm r}|^4
\end{eqnarray} 
	where the quantities indicated by suffices r are renormalized ones,
	and $Z$'s are the renormalization constants.
We can see that if 
\begin{eqnarray} 
	Z_\phi =Z_\lambda =0,\label{cc}
\end{eqnarray} 
	the Lagrangians ${\cal L}'_{\rm NJL}$ and ${\cal L}_{\rm Yukawa}$ 
	coincide, where we identify $\psi $, $\phi $ and $f$ 
	with 
\begin{eqnarray} 
  \psi =\sqrt {Z_\psi }\psi _{\rm r},\ \ \ 
  \phi ={Z_g \over Z_\psi }g_{\rm r}\phi _{\rm r}, \ \ \ 
  f={Z_g^2g^2_{\rm r}\over Z_\psi ^2Z_\mu  \mu _{\rm r}^2}
\end{eqnarray} 
	in the Yukawa model.
This condition (\ref{cc})
	is called the compositeness condition \cite{cc}.
Thus this Lagrangian for NJL model is the special case 
	of the renormalized Yukawa model 
	specified by the compositeness condition.
The compositeness condition gives rise to relations
	among coupling constants $g_{\rm r}$, $\lambda _{\rm r}$, 
	and the cut off $\Lambda $.
If the chiral symmetry is spontaneously broken,
	they imply relations among the fermion mass $m_f$, 
	the Higgs-scalar mass $M_H$, and the cutoff $\Lambda $.
Thus we can analyze everything in the NJL model
	by investigating the well-understood Yukawa model,
	and by imposing the compositeness condition 
	on the coupling constants and masses. 
Then what is urgent is 
	to work out the compositeness condition,
	and solve it for the coupling constants.

\noindent {\bf 2.2 Lowest order}

For an illustration, we begin with 
	the lowest-order contributions in $1/N$ expansion.
In the Yukawa model, 
	the boson self-energy part and the four-boson vertex part 
	are given by the diagrams 
\begin{eqnarray} 
\thicklines
\put(-150,20){
  \put( 50,3){  \Floop }
  \put( 85,0){  $\sim g_{\rm r}^2NIp^2$, }
  \put(175,3){  \hdots+(-18,0){13} 
                  \put(-5,-5){\line(1, 1){10}}
                  \put(-5, 5){\line(1,-1){10}}  }
  \put(200,0){  $=(Z_\phi -1)p^2$,}
}
\put(-150,-20){
  \put( 50,3){  \FloopX }
  \put( 85,0){  $\sim g_{\rm r}^4NI$,}
  \put(175,3){  \Xdots(0,7)
                  \put(0,-7){\line(0,1){14}}
                  \put(-7,0){\line(1,0){14}}  }
  \put(200,0){  $=(Z_\lambda -1)\lambda _{\rm r}$,}
}					\label{1loop}
\end{eqnarray} 
	where \hskip 0pt\put(5,3){\line(1,0){21}}\hskip 31pt 
	and \hdots+(5,3)8\hskip 31pt 
	are the fermion and boson propagator, respectively, and 
	$I$ is the divergent integral 
\begin{eqnarray} 
I=\cases{
     {\displaystyle {1\over 16\pi ^2}{1\over \epsilon }}
	             \ \  {\rm (dimensional\ regularization)}
		\ \left( \displaystyle \epsilon ={4-d\over 2}\right) \cr 
      \displaystyle {1\over 16\pi ^2}\log\Lambda ^2
                     \ \ \  {\rm (Pauli\ Villars\ regularization)} }
\end{eqnarray} 	
The renormalization constants $Z_\phi $ and $Z_\lambda $ should be chosen as 
\begin{eqnarray} 
  Z_\phi =1-g_{\rm r}^2NI,\ \ \ \ 
  Z_\lambda \lambda _{\rm r}=\lambda _{\rm r} - g_{\rm r}^4NI
\end{eqnarray} 
	so as to cancel out all the divergences in (\ref{1loop}).
Then the compositeness condition is obtained by putting 
	$Z_\phi $ and $Z_\lambda $ vanishing,
\begin{eqnarray} 
    0=1-g_{\rm r}^2NI,\ \ \ \ 0=\lambda _{\rm r} - g_{\rm r}^4NI,
\end{eqnarray} 
	and it is easily solved 
	to give the expressions for the coupling constants \cite{ES}.
\begin{eqnarray} 
    g_{\rm r}^2={1\over NI},\ \ \ \ \
    \lambda _{\rm r}=\displaystyle {1\over NI}.
\end{eqnarray} 
If the chiral symmetry is spontaneously broken
	the masses of the physical fermion and physical Higgs scalar
	are given in terms of cutoff: 
\begin{eqnarray} 
    m_f=g_{\rm r}\langle \phi \rangle 
	=\langle \phi \rangle /\sqrt {NI},\ \ \ \
    M_H=2\sqrt \lambda _{\rm r}\langle \phi \rangle 
	=2\langle \phi \rangle /\sqrt {NI}
\end{eqnarray} 
The Higgs mass is twice the fermion mass.
\begin{eqnarray} 
    2m_f=M_H
\end{eqnarray} 
These reproduce the well known results of the lowest order 
	Nambu-Jona-Lasinio model \cite{NJL}.

\noindent {\bf 2.3 Next-to-leading order}

Now we turn to the next-to-leading order in $1/N$ \cite{A}.
In the Yukawa model, 
	the boson self-energy part is given by the diagram
\begin{eqnarray} 
\put(-160,0){ \Floop\bigcirc (-9,6,3)\bigcirc (-5,3,3)\bigcirc (0,2,3)
                    \bigcirc (5,3,3)\bigcirc (9,6,3) }
\put(-120,0){ + the counter terms for all the sub-diagram divergences,}
\label{se2}
\end{eqnarray} 
\vskip -3mm\noindent 
where 
\put(0,0){
  \def\bub#1{{\thicklines\bigcirc (#1,3,10)}}
  \put(0,0){ \hcircles(10,3)5}
  \put(40,0){= \hdots+(5,3)9}
  \put(84,0){+}
  \put(95,0){\hdots+(2,3)3\bub{15}\hdots+(22,3)3}
  \put(130,0){+}
  \put(141,0){\hdots+(2,3)3\bub{15}\hdots+(22,3)3\bub{35}\hdots+(42,3)3}
  \put(195,0){+ $\cdots $.}
}\hskip 83mm
The renormalization constant $Z_\phi $ is calculated to be 
\vskip -7mm
\begin{eqnarray} 
	Z_\phi =1-g_{\rm r}^2NI-g_{\rm r}^2I-{1\over N}(1-g_{\rm r}^2NI)
	\log(1-g_{\rm r}^2NI)
\end{eqnarray} 
	so as to cancel out all the divergences in (\ref{se2}). 
The logarithm arises from the infinite sum 
	over the fermion loop insertions into the internal boson lines.
The four boson vertex part is given by the diagrams
\begin{eqnarray} 
\put(-120,0){  \FloopX\bigcirc (-6,9,3) \bigcirc (-5,4,3) \bigcirc (0,2,3) 
                      \bigcirc (5,4,3) \bigcirc (6,9,3) }
\put( -40,0){  \BloopVVX} 
\put(  40,0){  \BloopVLX}
\put( 120,0){  \BloopLLX}	\label{vt2}
\end{eqnarray} 
+ counter terms for the sub-diagram divergences.
The renormalization constant $Z_\lambda $ is calculated to be 
\vskip -10mm
\begin{eqnarray} 
Z_\lambda \lambda _{\rm r}&=&\lambda _{\rm r} - g_{\rm r}^4NI+8g_{\rm r}^4I
       +{20(\lambda _{\rm r} - g_{\rm r}^2)^2I\over 1-g_{\rm r}^2NI}
\cr &&
 -{1\over N}\left[ 2g_{\rm r}^2(1-g_{\rm r}^2NI)+20(\lambda _{\rm r} 
                  - g_{\rm r}^2)\right] 
\log(1-g_{\rm r}^2NI)
\end{eqnarray} 
	so as to cancel out all the divergences in (\ref{vt2}). 
The compositeness condition is given 
	by putting these expressions vanishing.
Though it looks somewhat complex at first sight,
	it can be solved by iteration to give 
	the very simple solution:
\begin{eqnarray} 
g_{\rm r}^2
	=\displaystyle {1\over NI}\left[ 1-{1\over N}
                  +O({1\over N^2})\right] ,
\ \ \ \ 
\lambda _{\rm r} 
	=\displaystyle {1\over NI}\left[ 1-{10\over N}+
               O({1\over N^2})\right] .
\end{eqnarray} 
The next-to-leading correction to the ratio of $M_H$ and $m_f$,
	which was 2 in the lowest order,
	is calculated to be:
\begin{eqnarray} 
	{M_H\over m_f}
	={g_{\rm r}\over \sqrt {\lambda _{\rm r}}}
	=2\left[ 1-{9\over 2N}+O({1\over N^2})\right] 
\end{eqnarray} 
For the case of $N=3$ of the practical interest,
	the corrections turn out to be too large,
	and the coupling constant $\lambda $ is negative,
	which implies that the Higgs potential is unstable.

\vskip 3mm\noindent 
{\large\bf 3. Induced Gauge Theory  --- Abelian ---}

We can apply \cite{AH} this method to the induced gauge theory \cite{ind},
	namely, the gauge theory with a composite gauge field.
It is given by the strong coupling limit 
	$f\rightarrow \infty $ \cite{Birula} of the
	vector-type four Fermi interaction model 
	for the fermion $\psi $ with the mass $m$:
\begin{eqnarray} 
   {\cal L}_{\rm 4F}
	=\overline \psi _j \left( i\slash \partial -m\right) \psi  
	-f\left( \overline \psi \gamma _\mu \psi \right) ^2,
\end{eqnarray} 
where $f$ is the coupling constant.
The Lagrangian ${\cal L}_{\rm 4F}$ is equivalent to 
\begin{eqnarray} 
   {\cal L}'_{\rm 4F}
   =\overline \psi 
	\left( i\slash \partial -m- \slash \backabit A\right) \psi 
\end{eqnarray} 
written in terms of the vectorial auxiliary field $A_\mu $.
Then we can see that 
	this is the special case of the renormalized gauge theory
\begin{eqnarray} 
   {\cal L}_{\rm G}
	=\overline \psi _{\rm r}\left( i Z_2 \slash \partial -Z_m m_{\rm r} 
	-Z_1 e_{\rm r}\slash \backabit A_{\rm r}\right) \psi _{\rm r}
	 -{ 1 \over 4 } Z_3 \left(   F_{\rm r}^{\mu \nu } \right)  ^2, 
\end{eqnarray} 
	specified by the compositeness condition
\begin{eqnarray} 
	Z_3=0,
\end{eqnarray} 
	where the quantities indicated by suffices r are renormalized ones,
	$e$ is the effective coupling constant,
	$Z$'s are the renormalization constants,
	and $F_{\rm r}^{\mu \nu }$ is the field strength of $A_{{\rm r}\mu }$.

The self-energy part of the gauge boson 
	is given by the diagrams 
\begin{eqnarray} 
  \put(-60,0){ \Floop }
  \put(  0,0){ \Floop \bigcirc (-9,6,3)\bigcirc (-5,3,3)\bigcirc (0,2,3)
	              \bigcirc (5,3,3)\bigcirc (9,6,3)}
  \put( 60,0){ \Floop \bigcirc (0,-10,3)\bigcirc (0,-5,3)\bigcirc (0,0,3)
	              \bigcirc (0,5,3)\bigcirc (0,10,3)}
\label{cgb}
\end{eqnarray} 
at the leading and the next-to-leading order.
The renormalization constant $Z_3$ is chosen 
	so as to cancel out the divergences in these diagrams.
After a lengthy calculation we obtain the following expression for $Z_3$:
\begin{eqnarray} 
Z_3=1-{e_{\rm r}^2 N \over 12\pi ^2 \epsilon  } 
    -{3e_{\rm r}^2 \over 16\pi ^2 }
	\left[  1+\left(  1-{12\pi ^2\epsilon \over e_{\rm r}^2 N }\right)  
	\ln\left(  1-{e_{\rm r}^2 N \over 12\pi ^2\epsilon }\right) \right] ,
\end{eqnarray} 
where $\epsilon =(4-d)/2$ with the dimension $d$.
Then the compositeness condition $Z_3=0$ 
	is solved to give the simple solution:
\begin{eqnarray} 
	e_{\rm r}^2 = { 12 \pi ^2 \epsilon \over N }
	\left[  1-{9\epsilon \over 4 N }\right] .
\end{eqnarray} 
The correction term $9\epsilon /4N$ is 
	naturally suppressed by the small factor $\epsilon $.
It justifies the lowest order approximation of this model
	unlike in the case of the 
	aforementioned NJL model of the scalar composite.
The origin of the suppression factor is traced back to 
	the gauge cancellation of the leading divergence 
	in the next-to-leading (in $1/N$) diagrams in (\ref{cgb}).

So far we assumed that all the fermions has the same charges for simplicity.
If the charges $Q_i$ are different, the expression is modified as
\begin{eqnarray} 
	e_{\rm r}^2 ={ 12 \pi ^2 \epsilon \over \sum _j Q_j^2 }
	\left[  1-{9\epsilon \sum _j Q_j^4\over 4(\sum _j Q_j^2)^2}\right] .
\end{eqnarray} 
If we apply this to the quantum electrodynamics
	with 3 generations of quarks and leptons,
	$\epsilon $ is estimated to be 6$\times 10^{-3}$,
	which implies the next-to-leading order correction amounts 
	only to 0.1\% of the lowest order term.

\vskip 3mm\noindent 
{\large\bf 4. Induced Gauge Theory  --- Nonabelian ---}

Now we apply it \cite{AH2} to the non-abelian induced gauge theory.
This is the new result obtained after my talk at SCGT96, 
International Workshop on Perspectives of Strong Coupling Gauge Theories, 
held in Nagoya, Japan, in November 1996 \cite{SCGT}. 
We start with the four Fermi Lagrangian  
\begin{eqnarray} 
   {\cal L}_{\rm 4F}
	=\overline \psi \left( i\slash \partial -m\right) \psi  
	-f\left( \overline \psi \lambda ^a\gamma _\mu \psi \right) ^2
\end{eqnarray} 
for the fermion $\psi $ with $N_c$ gauged color and $N_f$ ungauged flavor,
	where $\lambda _a$ ($a=1,\cdots ,N_c^2-1$) 
	is the $N_c\times N_c$ Gell-Mann matrix,
$m$ is the mass of the fermion and $f$ is the coupling constant.
In the strong coupling limit $f\rightarrow \infty $, 
	the Lagrangian ${\cal L}_{\rm 4F}$ is equivalent to the linearized one 
\begin{eqnarray} 
    {\cal L}'_{\rm 4F}
	=\overline \psi 
	\left( i\slash \partial -m-\lambda _a\slash \backabit A^a\right) 
	\psi 
\end{eqnarray} 
written in terms of the auxiliary field $A_\mu ^a$.
Then ${\cal L}'_{\rm 4F}$ is the special case of the renormalized gauge theory
\begin{eqnarray} 
{\cal L}_{\rm G}&=&
	\overline \psi _{\rm r}
	\left( i Z_2 \slash \partial 
	    -Z_m m_{\rm r}
	    -Z_1 g_{\rm r}\lambda ^a\slash \backabit A_{\rm r}^a\right) 
	\psi _{\rm r}
\cr &&
  -{ 1 \over 4 } Z_3 \left( \partial _\mu  A_{{\rm r}\nu }^a 
	- \partial _\nu  A_{{\rm r}\mu }^a 
        + Z_3^{1/2} Z_g g_{\rm r}f^{abc} A_{{\rm r}\mu }^b A_{{\rm r}\nu }^c
 	\right)  ^2 ,
\end{eqnarray} 
	specified by the compositeness condition 
\begin{eqnarray} 
	Z_3=0,
\end{eqnarray} 
	where the quantities indicated by suffices r are renormalized ones,
	$g$ is the effective coupling constant,
	$Z$'s are the renormalization constants,
	and $F_{{\rm r}\mu \nu }^{a}$ 
	is the field strength of $A_{{\rm r}\mu }^a$.
The self-energy part of the gauge boson, 
	at the leading order and the next-to-leading order in $1/N_f$,
	is given by the diagrams 
\begin{eqnarray} 
  \put(-180,0){ \Floop }
  \put(-120,0){ \Floop \bigcirc (-9,6,3)\bigcirc (-5,3,3)\bigcirc (0,2,3)
		       \bigcirc (5,3,3)\bigcirc (9,6,3)}
  \put(-60,0){ \Floop  \bigcirc (0,-10,3)\bigcirc (0,-5,3)\bigcirc (0,0,3)
		       \bigcirc (0,5,3)\bigcirc (0,10,3)}
  \put(  0,0){ \BloopVV}
  \put(60,0){ \BloopVL}
  \put(120,0){ \BloopLL} 
  \put(180,0){ \FPloop }		\label{naloop}
\end{eqnarray} 
	where \fcircle (5,3,1)\fcircle (10,3,1)\fcircle (15,3,1)
              \fcircle (20,3,1)\fcircle (25,3,1)\hskip 30pt
	is the Fadeev Popov ghost propagator.
The renormalization constant $Z_3$ is chosen 
	so as to cancel out the divergences in the diagrams in (\ref{naloop}).
After a lengthy calculation we obtain the expression:
\begin{eqnarray} 
Z_3 &=& 1 - { 2\over 3} N_f g_{\rm r}^2 I
         + {11\over 3} N_c g_{\rm r}^2 I
         - {\alpha \over 2} N_c g_{\rm r}^2 I 
             (1 - { 2\over 3} N_f g_{\rm r}^2 I)
\cr &&
   + {3\over 2} N_c ( {3\over 2N_f} 
      - g_{\rm r}^2 I)\ln(1 - { 2\over 3} N_f g_{\rm r}^2 I)
    +O({1\over N_f^3}).
\end{eqnarray} 
Then the compositeness condition $Z_3=0$ 
	is solved to give the simple solution:
\begin{eqnarray} 
	g_{\rm r}^2 = \displaystyle {3\over 2N_fI}
            \left[ 1+{11N_c\over 2N_f}+O({1\over N_f^2})\right] .
\end{eqnarray}

This results exhibit several interesting features.
\\1) The gauge boson self-energy part is purely transversal
	(i.e.\ proportional to $(p^2g_{\mu \nu }-p_\mu p_\nu )$)
	at all order in $g_{\rm r}$ at the next-to-leading order.
In the practical calculation, 
	this is realized, at the lowest order in $g_{\rm r}$, 
 	by adding the Fadeev Popov loop, 
	and in the higher order, 
	by cancellations of the non-transversal parts.
\\2) Though the $Z_3$ itself does depend on gauge parameter $\alpha $,
	the solution $g_{\rm r}$ to the compositeness condition $Z_3=0$ 
   	does not.
This should be so 
	because the coupling constants and the compositeness scale
	are observable object.
\\3) The condition that the next-to-leading contribution
	should not exceed the leading contribution
	implies that 
\begin{eqnarray} 
	N_f>{11N_c\over 2}.
\end{eqnarray} 
Note that this critical value for $N_f$ and $N_c$
	coincides with that for asymptotic freedom.
When the gauge theory is asymptotically free,
	the next-to-leading contributions are too large,
	so that the gauge bosons cannot be a composite of the above type.
And when it is asymptotically non-free, 
	the next-to-leading order contributions are reasonably suppressed,
	and the gauge boson can be interpreted as a composite.

\vskip 3mm\noindent 
{\large\bf 5. Summary}

In summary, by solving the compositeness condition, 
	under which the Yukawa-type model coincides the NJL type model,
	we obtained the expressions for the effective coupling constants
	in terms of the compositeness scale 
	at the next-to-leading order in $1/N$.
In the NJL model with a scalar composite,
	the next to leading contribution to $g_{\rm r}^2$ is $-1/N$,
	and that to $\lambda _{\rm r}$ is $-10/N$, 
	which are too large for $N_c=3$ of our practical interest,
	and imply unstable Higgs potential.
On the other hand,
	in the induced gauge theory with abelian gauge symmetry,
	the correction term to $e_{\rm r}^2$ is $-9\epsilon /4N$, 
	which is reasonably suppressed by the small factor $\epsilon $,
	and amounts to only 0.1\% when it is applied to the QED.
In the non-abelian gauge theory,
	the correction term to $g_{\rm r}^2$ is $11N_c/2N_f$.
This implies that, when the gauge theory is asymptotically free, 
	the next to leading contribution is too large,
	and we can not take the gauge boson as a composite of this type.
On the other hand, when it is asymptotically non-free,
	the corrections are suppressed, 
	and the gauge boson is safely taken as a composite.
We expect that these methods and results will be useful
	in disclosing the nature of composite objects
	in particle and nuclear physics in the future.

\noindent {\bf Acknowledgment}

The author would like to thank Professors H.~Terazawa and T.~Matsuki 
for stimulating discussions, and Professor T.~Hattori for collaborations.
He wishes also to thank 
Professor Y.~Koide and the members of the organizing committee of this 
workshop 
for their kind hospitalities extended on him during his stay in Shizuoka
and for arranging the opportunity for him to present this work
in the workshop.

\end{document}